\documentclass[12pt]{iopart}

\usepackage{graphicx}
\usepackage{subfigure}

\begin{document}

\title[Modelling the Self-Assembly of Virus Capsids]{Modelling the Self-Assembly of Virus Capsids}

\author{Iain G. Johnston$^1$, Ard A. Louis$^1$ and Jonathan P. K. Doye$^2$}

\address{$^1$Rudolf Peierls Centre for Theoretical Physics, 1 Keble Road, Oxford OX1 3NP, United Kingdom, $^2$ Physical \& Theoretical Chemistry Laboratory, University of Oxford, South Parks Road, Oxford OX1 3QZ, UK}

\begin{abstract}
We use  computer simulations to study a model, first proposed by Wales \cite{wales2005energy}, for the reversible and monodisperse self-assembly of simple icosahedral virus capsid structures.   The success and efficiency of assembly as a function of thermodynamic and geometric factors can be qualitatively related to the potential energy landscape structure of the assembling system. Even though the model is strongly coarse-grained, it exhibits a number of features also observed in experiments, such as sigmoidal assembly dynamics, hysteresis in capsid formation and numerous kinetic traps.  We also investigate the effect of macromolecular crowding on the assembly dynamics.  Crowding agents generally reduce capsid yields at optimal conditions for non-crowded assembly, but may increase yields for parameter regimes away from the optimum.  Finally, we generalize the model to a larger triangulation number $T=3$, and observe more complex assembly dynamics than that seen for the original $T=1$ model.
\end{abstract}

\pacs{81.16.Dn,87.15.ak,87.15.km,81.16.Fg,87.15.A-}
%81.16.Dn       Self-assembly
%87.15.ak       Monte Carlo simulations
%87.15.km       Protein-protein interactions
%81.16.Fg       Supramolecular and biochemical assembly
%87.15.A-       Theory, modeling, and computer simulation

\section{Introduction}
One of the simplest examples of self-assembly in biology is that of the virus capsid. The protective protein coat surrounding the viral genetic material is assembled into its monodisperse form reversibly from a large number of quasi-identical subunits, or capsomers.   In some cases, this can even occur \emph{in vitro}, as famously first demonstrated for the tobacco mosaic virus where the capsids dissociated upon raising the pH of a solution, but reversibly reassembled into complete capsid structures upon the subsequent lowering the pH  back to the initial conditions \cite{fraenkel1955reconstitution}.    Although in nature the assembly of virus capsids can be more complex, with nucleic acids, scaffolding proteins, and other constituents playing a role, there are a good  number of  viruses where successful reversible \emph{in vitro} assembly can occur from just the purified proteins. Well studied examples include icosahedral viruses such as the cowpea chlorotic mottle virus (CCMV) \cite{deng2008cryo}, the hepatitis B virus (HBV) \cite{zlotnick2000mechanism} and the human papillomavirus (HPV) \cite{zlotnick1999theoretical}.    These experiments suggest that it is possible to encode all the necessary assembly information into the individual capsomers themselves.  As such, virus self-assembly \emph{in vitro} is a  paradigmatic example of monodisperse self-assembly. 
 A better understanding of the underlying physics involved will not only lead to new biological insights, but may also stimulate novel applications in nanotechnology.
 
 Viruses  typically vary in size from about 2 to 200$\,$nm in diameter and are extremely successful organisms.  They can be  found in a wide variety of environments and it has recently been estimated that they constitute a larger fraction of the total biomass on earth than eukaryotes (the family that includes animals and plants) \cite{suttle2005viruses}.   About half of all virus families have icosahedral symmetry. As first pointed out by Caspar and Klug \cite{caspar1962cold}, the structure of such virus capsids can be understood using the concept of quasi-equivalence.  That is, identical proteins can occupy different location types within the capsid structure. Some capsomers group around an axis of five-fold symmetry, and others around axes of six-fold symmetry, occupying geometrically distinct locations despite their structural similarity \cite{willits2003effects,hanslip2006assembly,fane2003mechanism}.  A simple geometric argument suggests that capsids can be made up of  $12$ pentagonal and $10(T - 1)$ hexagonal structural units, where the triangulation number $T$ is restricted to numbers that can be expressed by $T = a^2 + ab + b^2$, with $a$ and $b$ non-negative integers (e.g. $T=2$ is not possible, but $T=1$ and $T=3$ are). The smallest $T = 1$ viruses are made up of  just 12 pentameric units.  Many of these are satellite viruses that are parasitical to larger viruses, but others such as the alfafa mosaic virus \cite{carrillo2009viper} are viable on their own.   More complex viruses have higher triangulation numbers, including CCMV ($T=3$), HBV ($T = 4$) and HPV ($T = 7$) \cite{carrillo2009viper}, and the largest to date is estimated to be an enormous $T \simeq 1000$  for the giant Mimivirus \cite{xiao2009structural}.  
 
Experimental studies have observed the \emph{in vitro} assembly of virus capsids using light scattering \cite{casini2004vitro,mukherjee2008quantitative} and electron microscopy \cite{casini2004vitro,sorger1986structure}. Several key features are commonly observed.  Firstly, the yield of complete capsids is found to be a sigmoidal function of time, with an initial lag period during which no capsids are produced, followed by a period of rapid completion before the yield reaches a plateau. Secondly, under fast growth conditions such as high subunit concentrations or strong attractions, assembly is seen to be vulnerable to a kinetic trap in which an excessive number of nuclei are formed. Each of these nuclei grow, consuming subunit monomers in the process and leading to a rapid depletion of the monomer population (``monomer starvation'') and the eventual production of a large number of part-formed capsids. Also observed is the formation of ``monster particles'', consisting of malformed clusters of part-formed capsids. Thirdly, hysteresis is also commonly found where fully formed capsids remain stable in conditions under which they would not form from individual capsomers.

Modeling virus assembly at the atomistic level is prohibitively expensive.  In a recent study, a fully atomistic simulation of a complete $T=1$ icosahedral satellite tobacco mosaic was performed \cite{freddolino2006molecular}.  But even this simulation involved over a million individual atoms (mostly solvent), and so could only sample about 50 nanoseconds of time, whereas dynamic assembly occurs on much longer time-scales (up to seconds).      
 
Many studies of virus assembly have thus, by necessity, treated the process by using strongly coarse-grained models.  Important early work was done by Zlotnick and co-workers \cite{zlotnick1999theoretical,zlotnick1994build}, who employed kinetic equations that measure the flux between populations of different sized assemblies.  The spatial location of the different species is averaged over, and in most of the work,  the sets of possible reactions and intermediates was limited to the addition of single capsid proteins to their equilibrium position in capsomers, thus ignoring many potential intermediate states.  Nevertheless, these kinetic studies reproduced important experimentally observed features such as monomer starvation and hysteresis, and showed that relatively weak association energies are sufficient for capsid assembly.    The question of how important the approximation of  neglecting all but the most stable intermediates in the assembly pathway is still under active debate \cite{schwartz2005local, nguyen2007deciphering, misra2008pathway}.

Direct simulations of model coarse-grained protein assemblies have the advantage that spatial fluctuations and a much wider range of intermediate states are naturally included.  Some of the first were
performed by Rapaport and collaborators \cite{rapaport1999supramolecular,rapaport2004self} who used molecular dynamics (MD)  simulations of triangular and trapezoidal units that assemble into capsids.  However, these simulations suffered from drawbacks such as not always satisfying detailed balance and the use of unrealistic ballistic dynamics.  In a recent paper, Rapaport  performed fully reversible simulations where complete icosahedra assembled from 20 triangular particles in a background fluid\cite{rapaport2008role}.  There is now a considerable body of work using MD, Brownian dynamics or Monte Carlo (MC) simulations of coarse-grained particles to model the self-assembly of viruses \cite{nguyen2007deciphering,hagan2006dynamic,elrad2009mechanisms,chen2007simulation,wilber2009monodisperse,nguyen2009invariant} and other objects that self-assemble into monodisperse clusters \cite{wilber2007reversible,glotzer2004materials,van2006symmetry,glotzer2007anisotropy,wilber2009self}.
These studies exhibit a number of similar trends.  For example, to achieve self-assembly, the temperature must be low enough that the target structure is thermodynamically stable, but not so low that incorrectly bonded particles cannot separate.  Similarly, the design parameters must be specific enough to favour the target structure over alternative structures, but not so constrained as to hamper kinetic accessibility of the desired structure \cite{wilber2007reversible}.     Given that these basic trends are seen by such a wide variety of models, it suggests that they are robust features of self-assembly that will be relevant to the physical case of virus self-assembly {\em in vitro}.

A different way of approaching the design problem for virus capsids is to study the potential energy surface (PES), a high dimensional function that describes how the potential energy depends on the coordinates of the $N$ particles in a system.  It is closely related to the concept of a  free-energy landscape, which also includes the effects of temperature \cite{wales2003energy}.  This point of view has been particularly explored in the context of protein folding, which can also be viewed as a self-assembly phenomenon, where the particles (the amino acids) are all connected together instead of being free as they are in virus self-assembly.    An important concept for protein folding is the idea of a ``folding funnel'' \cite{wales2003energy,leopold1992protein,bryngelson1989intermediates,dobson2003protein},  that helps explain how a protein can overcome the Levinthal paradox \cite{levinthal1969fold}, which states that it is impossible for a protein to find its folded state on a physical time-scale by a completely blind search because the number of states accessible to a typical protein is astronomically large.   Instead,  the ``funnel'' topography of the PES  helps guide the system through a directed search towards the free-energy minimum.  
  By analogy, one might expect that the energy landscape of a self-assembling system must also show a funnel-like topology if the system is to assemble.  Indeed, we have recently calculated free-energy landscapes of model patchy particles, and shown that this feature can help rationalize the dependence of self-assembly yields on design parameters \cite{wilber2009monodisperse,wilber2009self}.

In a recent paper, Wales \cite{wales2005energy} proposed a model for the assembly of $T=1$ viruses based on a set of 12 pentagonal units, which has since been recently extended \cite{fejer2009energy} to model $T=3$ capsids. In Ref. \cite{wales2005energy}, the PES for a single connected 12-mer was characterised, and stationary states were used to analyze the topology of the landscape for a number of different design parameters. Rugged, glassy landscapes and landscapes with insufficient funnelling were proposed to hinder assembly. Those landscapes that exhibit a more funnel-like topography were predicted to promote assembly. However, there are a few caveats to this picture.  Firstly, the PES  was calculated for a connected 12-mer, which on its own does not explain why 12-mers form rather than clusters of other sizes. In Ref. \cite{fejer2009energy} the PES for 24 particles, using their extended model, was shown to have separate funnels, one corresponding to two distinct icosahedral capsids and the other to a single 24-particle cluster, the former being more stable. More generally, the PES is a very high dimensional function, but the analysis of its topology is usually  done by projecting it down to a much lower dimensional representation.  It is therefore interesting to see how robust these PES based predictions are.  

To investigate the PES predictions, we perform MC computer simulations on the same model and study its assembly dynamics.  We find good agreement with Wales' predictions, giving further evidence that the energy landscape picture can help rationalize how nature achieves the design of individual particles that can self-assemble into well defined mono-disperse shapes.    Even though this model is strongly coarse-grained, we observe behaviour that is also seen in experiments, including sigmoidal assembly dynamics, hysteresis in capsid formation, and a variety of different kinetic traps.  

We also extend the model in two ways.  We first add crowding agents to model the fact that when viruses assemble {\em in vivo}, they do so in the densely packed environment of the cell.  The crowding lowers the assembly efficiency compared to the uncrowded case for parameters near the optimal assembly, but can increase efficiency in regions where the uncrowded virus assembly proceeds less well. 

 We next extend the model to $T=3$ viruses by including 20 hexagonal units per capsid, and find a region of parameter space where assembly is successful.   We also investigate a number of different kinetic traps and pathways to assembly not observed for the $T=1$ model.   In contrast to the simpler $T=1$ model,  the assembly efficiency drops considerably when more than one capsid is simulated, in part because the pentagons and the hexagons must come together in the right numbers per capsid.

We proceed as follows: In section 2 we describe the model and the MC simulation method we use, and in section 3 we discuss simulation results for the the $T=1$ assembly dynamics.  Section 4 includes the two extensions of our model, namely  crowded assembly and  the $T=3$ virus model.

\section{Methods}

\subsection{Model}

\begin{figure*}
\begin{center}
\includegraphics[width=8cm]{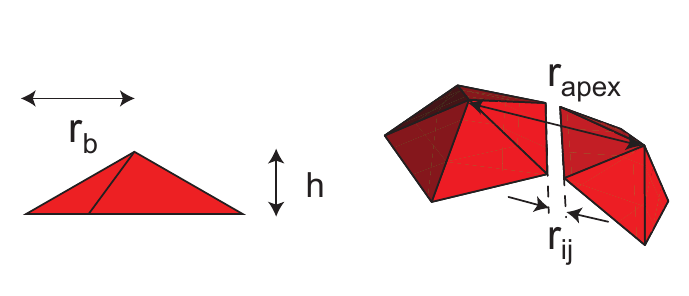}
\end{center}
\caption{The rigid pentagonal pyramid capsomer units have a height of $h$ and a distance between the base centre and the vertices of $r_b$.  Also shown here is the distance $r_{apex}$ between two apices and the distance $r_{ij}$ between two basal vertices on different capsomers.}
\label{capsomerfig}
\end{figure*}

\begin{figure*}
\includegraphics[width=16cm]{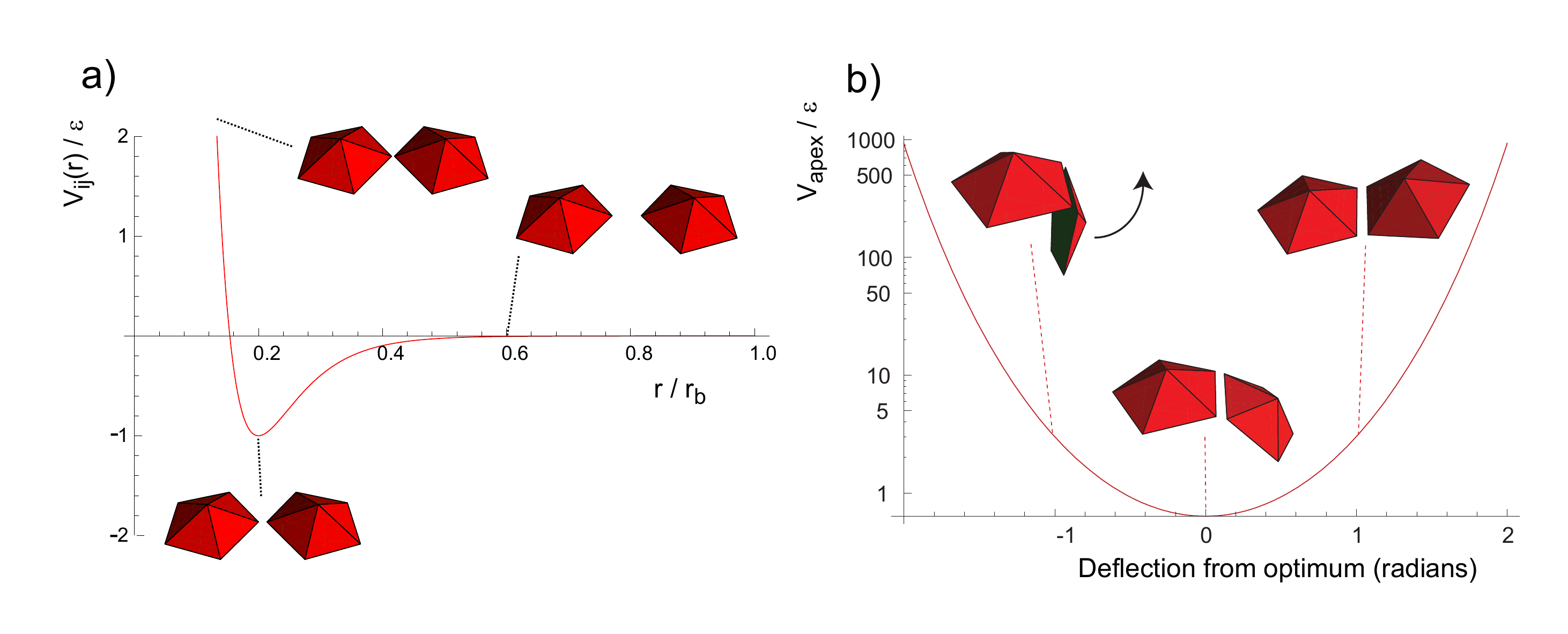}
\caption{(a) Morse interaction $V_{ij}(r)$ between base points on model capsomers. (b) The repulsive potential $V_{apex}(r)$ between capsomer apices helps set the curvature of the fully formed capsids. Note that capsomer illustrations are not to same scale as graphs.}
\label{fig1}
\end{figure*}

We use the same model as Wales \cite{wales2005energy}. The  capsomers are represented as rigid pentagonal pyramids with a distance $r_b$ from the centre of the capsomer base to  the basal vertices, and with an apex site at a height $h$ above the centre of the base, as illustrated in Fig.~\ref{capsomerfig}. 
Representing a capsomer $c_i$ by vectors corresponding to its base points, $\{ \mathbf{p_i^1}, ..., \mathbf{p_i^5} \}$, and its apex $\mathbf{a_i}$, the potential between two capsomers $c_i$ and $c_j$ is:
\begin{equation}
V(c_i, c_j) = V_{apex}\left( \left| \mathbf{a_i} - \mathbf{a_j} \right| \right) + \sum_{u=1}^5 \sum_{v=1}^5 V_M\left( \left| \mathbf{p_i^u} - \mathbf{p_j^v} \right| \right).
\end{equation}
where $V_M(r)$ is a Morse potential of the form:
\begin{equation}
V_M(r_{ij})  =  \epsilon \left( e^{\rho (1 - r_{ij}/r_e)} - 2 \right)e^{\rho (1 - r_{ij}/r_e)} 
\end{equation}
and $V_{apex}(r)$ is a purely repulsive interaction of the form:
\begin{equation}
V_{apex}(r)  =  \epsilon_{R} \left( \frac{\sigma}{r_{apex}} \right) ^{12}.
\end{equation}
Here $\epsilon$  defines the unit energy (from which reduced temperature $T^*$ is derived), and $r_b$, the distance between the centre of a capsomer's base and one of its vertices, defines unit distance. $r_e$, the length scale of the Morse potential, is set to $0.2\,r_b$. The range of the interaction is set to $\rho=0.6\,r_b$,  and the strength of the repulsive term is set to $\epsilon_{R}=\frac{\epsilon}{2}$. Capsomers can have variable height $h$, and unless  otherwise noted, $h$ is set to $r_b/2$. The apex-apex repulsion length-scale is set to $\sigma =2.10\,r_b$, the distance between two apices in a complete capsid when $h = 0.5\,r_b$ \cite{wales2005energy}, so that $V_{apex}(r) = \epsilon_R$ when the capsid is fully formed.
  The attractions between the vertices allow the capsomers to bond, and the repulsive interaction between the apex sites sets the curvature of the capsid, see Fig.~\ref{fig1} for an illustration of these potentials. 

\subsection{Simulations}

To perform the simulations we employed a standard Metropolis Monte Carlo scheme with only local translation and rotation moves of the individual capsomers.
 The small-scale and random nature of the steps applied generate diffusive motion similar to that expected for proteins in solution \cite{tiana2007use,kikuchi1991metropolis}.   The real dynamics of aggregating particles in solution are more complicated \cite{whitelam2007avoiding}, and hydrodynamic effects, which are neglected in Brownian dynamics, may also play a role \cite{padding2006hydrodynamic}.  However,  local move MC should be adequate to capture overall trends in a diffusion limited system.  Translation and rotation moves are chosen stochastically, each with a $50\%$ probability. Due to the coarse-grained and conceptual nature of this model, we did not attempt to  adjust move size and frequency to reflect quantitatively accurate ratios of rotational to translational diffusion. 

The simulations were performed in a cubic simulation box with periodic boundary conditions.  The number $N_5$ of pentagonal capsomers was varied between $12$ and $120$.  To initialize the system, 
 $10^4$ MC cycles (where a cycle is $N_5$ steps) were performed at $T^* = 10\, \epsilon k_B^{-1}$.  Assembly was then monitored for runs up to $5 \times 10^5$ MC cycles unless otherwise stated.

To analyze the clusters of bonded capsomers we employed the following protocol.  
 Two capsomers $c_i$ and $c_j$ are considered to be bonded if $V(c_i, c_j) < -\epsilon$.  Since the potential energy of two capsomers in a perfectly bonded configuration is $-2\epsilon + \epsilon_R$ (terms respectively from the base-point bonds and apex repulsion), equal to $\frac{3}{2}\epsilon$, our bonding definition allows a degree of thermal fluctuation while ensuring that capsomers are still bound. A cluster is then a set of capsomers where each capsomer in the set is reachable from any other by following a series of bonds. We define a  cluster size order parameter, $C$, as the size of the largest cluster present in a simulation as well as  a geometric capsid order parameter, $Q$, defined as:
\begin{equation}
Q = \frac{1}{N_5} b_5,
\end{equation}
where $b_5$ is the number of capsomers in the simulation bonded to exactly five other capsomers with bond angles consistent with that of a fully formed capsid structure. This geometric order parameter is useful to distinguish the correct structure from those with inverted capsomers, predicted by Wales \cite{wales2005energy} to represent significant kinetic traps in this model.

\section{Results for $T=1$ capsid assembly}

\subsection{Assembly yields with density and temperature}

\begin{figure}
\begin{center}
\includegraphics[width=8cm]{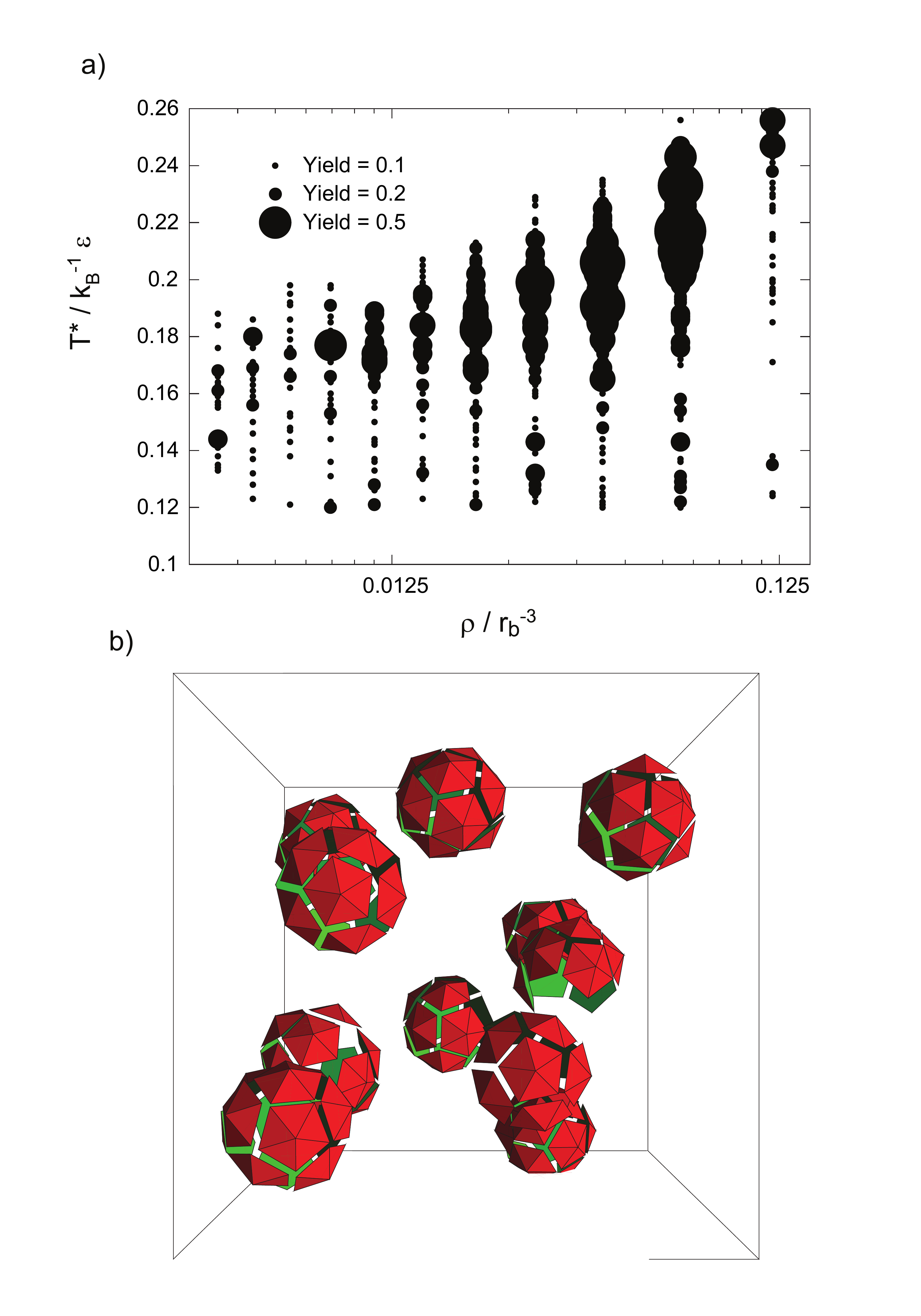}
\end{center}
\caption{(a) Assembly yield with number density $\rho$ and reduced temperature $T^*$. (b) Snapshot of capsid assembly at $T^* = 0.22\, \epsilon k_B^{-1}, \rho = 5.5 \times 10^{-2}\, r_b^{-3}$ (packing fraction $\phi \simeq 0.27$), showing a $60\%$ yield of capsids.  In this case there are 11 clusters, because a number of capsids are only partially formed and there are no more monomers left to complete the assembly (the monomer starvation trap).}
\label{densityfig}
\end{figure}

\begin{figure}
\begin{center}
\includegraphics[width=8cm]{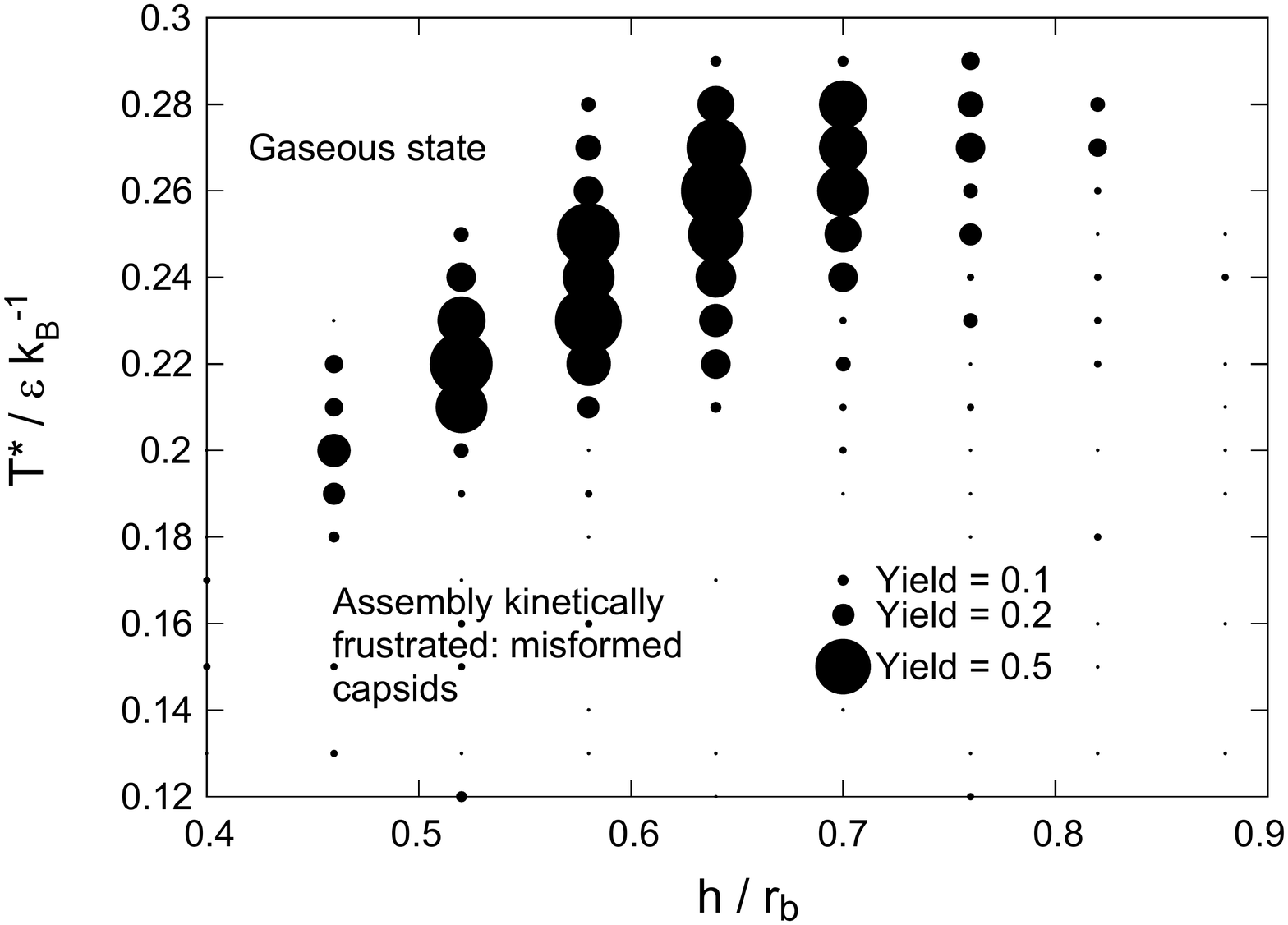}
\end{center}
\caption{Assembly yield with capsomer height $h$. As predicted by Wales \cite{wales2005energy}, 
the optimal assembly occurs at intermediate values of $h$ where the PES is predicted to show ``funnel''-like features.}
\label{heightfig1}
\end{figure}

For a fixed height $h = 0.5\,r_b$, we simulated $N_5 = 120$ capsomers and studied how the assembly yield, defined as the fraction of fully formed capsids (maximum of $10$), varies as a function of temperature and number density $\rho = N_5/L^3$, where $L$ is the length of one side of the cubic simulation box (periodic boundary conditions are applied).    The results are shown in Fig.~\ref{densityfig}.  We observe firstly that for a fixed density, the region of optimal assembly is bounded from above and below in temperature.  If the temperature drops too low, then the attractions are too strong and mis-bonded capsomers cannot dissociate and reassemble in correct configurations. If the temperature is too high, the attractions are not strong enough to ensure bonding, and the high entropy disordered state  is favoured.    These results mirror those found by a number of other investigators \cite{hagan2006dynamic,wilber2009monodisperse,wilber2007reversible}.

Similarly, for a fixed temperature, there is also a window of densities for which optimal yields are obtained.   As discussed for example in the work of Hagan and Chandler \cite{hagan2006dynamic}, this trend is due to a tradeoff between having many subunit collisions with increasing density, and avoiding kinetic traps that create amorphous structures at higher densities.     Indeed we find many partially formed shells at low densities, and amorphous bonded structures at higher densities.  The optimal assembly region is around  $\rho \simeq 5.5 \times 10^{-2}\, r_b^{-3}$, at $T^* \simeq 0.22\, \epsilon k_B^{-1}$. We define $\rho^* = 5.5 \times 10^{-2}\, r_b^{-3}$ for use in further simulations.
To express this as a packing fraction, we approximate the volume occupied by each capsomer as a sphere of diameter $\sigma$, the length scale of the repulsive interaction. Using this measure, $\rho^*$ corresponds to a  capsomer packing fraction of $ \phi = \pi \rho^* \sigma^3/6 \simeq 0.27$. 

The yield of complete capsids is observed to be sigmoidal with time, with a brief period of zero yield during which nucleation occurs, followed by a rise in yield as capsids beging to form, plateauing as monomers are depleted and no further assembly can occur.

\subsection{Assembly yield with capsomer height}

\begin{figure}
\begin{center}
\includegraphics[width=6cm]{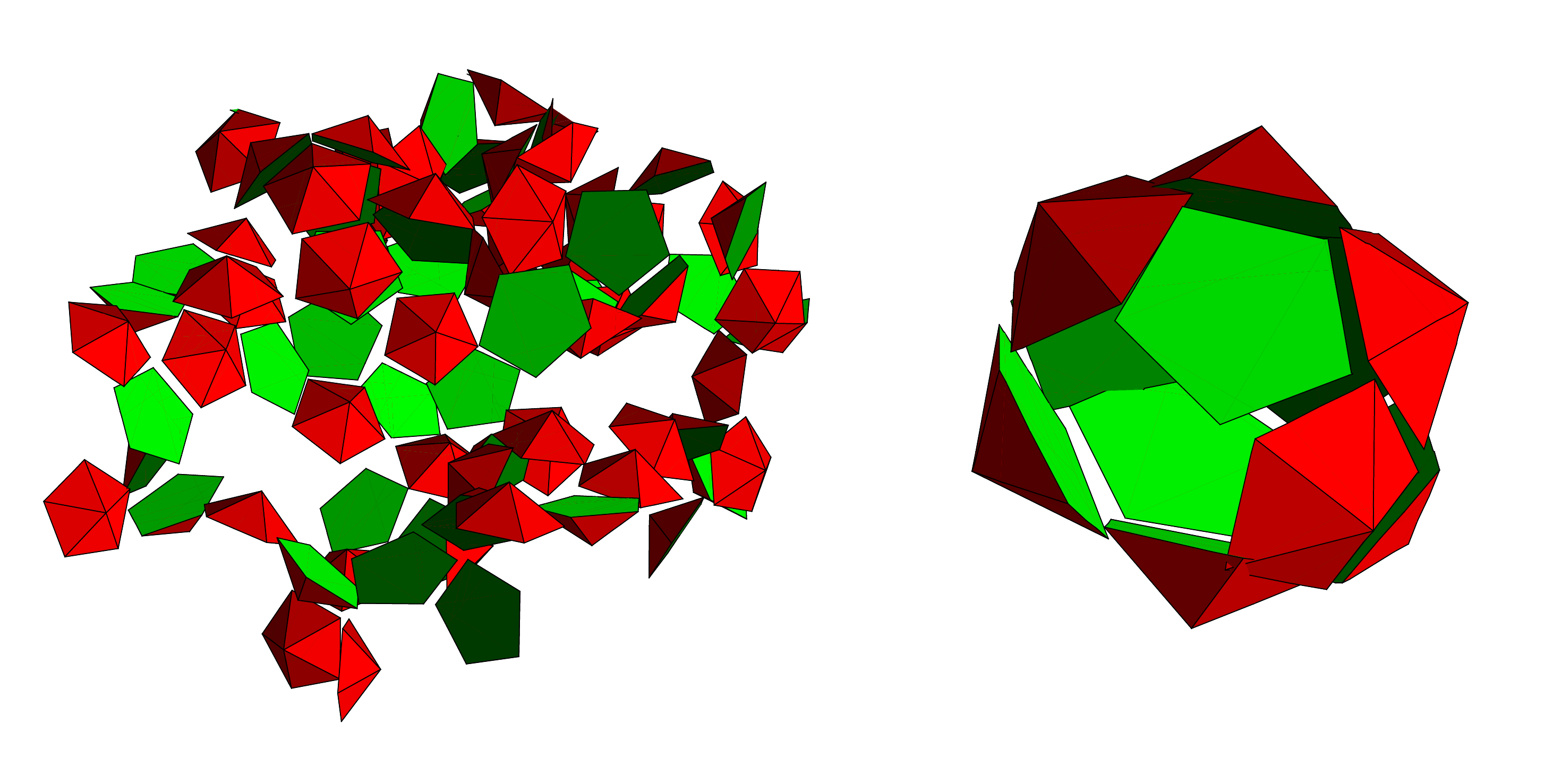}
\end{center}
\caption{Low-temperature kinetic traps in the formation of a $T=1$ capsid. (left) $h = 0.4\, r_b$: clustering into amorphous structure. (right) $h = 0.75\, r_b$: kinetic traps often involve one or more capsomers in the incorrect geometry, here inverted, as predicted by Wales \cite{wales2005energy}.}
\label{heightfig2}
\end{figure}

Wales analysed the PES for different values of the capsomer height $h$ \cite{wales2005energy}.
For $h=0.5\,r_b$, he observed a funnel-like topology that he predicted would facilitate assembly.  For a smaller capsomer height $h=0.35\, r_b$, the fully formed icosahedral capsid is still the global minimum, but the potential energy gradient and thus the driving force towards the formation of closed shells is diminished.  Furthermore, he predicted a kinetic trap made up of loosely packed capsomers.  Similarly, for larger capsomer height $h=0.75\,r_b$, there are competing low-lying energy states where capsomers join the capsid structure  with their apices pointing inwards.  Again, assembly is predicted to be hindered by these kinetic traps.

We simulated the system at a fixed density $\rho^*$, but for different capsomer heights and temperatures.
The results are shown in Figs.~\ref{heightfig1} and \ref{heightfig2}.  In agreement with Wales, we find that optimal assembly occurs at an intermediate $h$,  with an optimum closer to $h=0.6\,r_b$, and very low yields for $h < 0.35\,r_b$ or $h > 0.75\,r_b$.  As illustrated in Fig.~\ref{heightfig2},  for the larger $h$ we observe kinetic traps characterized by shells with inverted capsomers and    for low $h$ we find kinetic traps where the system forms an extended amorphous structure. 

\subsection{Assembly mechanisms and thermodynamics for a single capsid}

To study the assembly mechanism of a capsid in more detail, simulations were carried out for 12 capsomers at an effective of density $\rho^*$ and for different temperatures $T^*$.  
The cluster size order parameter $C$ and the geometric capsid order parameter $Q$ were averaged over ten independent simulations. 

\begin{figure*}
\begin{center}
\includegraphics[width=8cm]{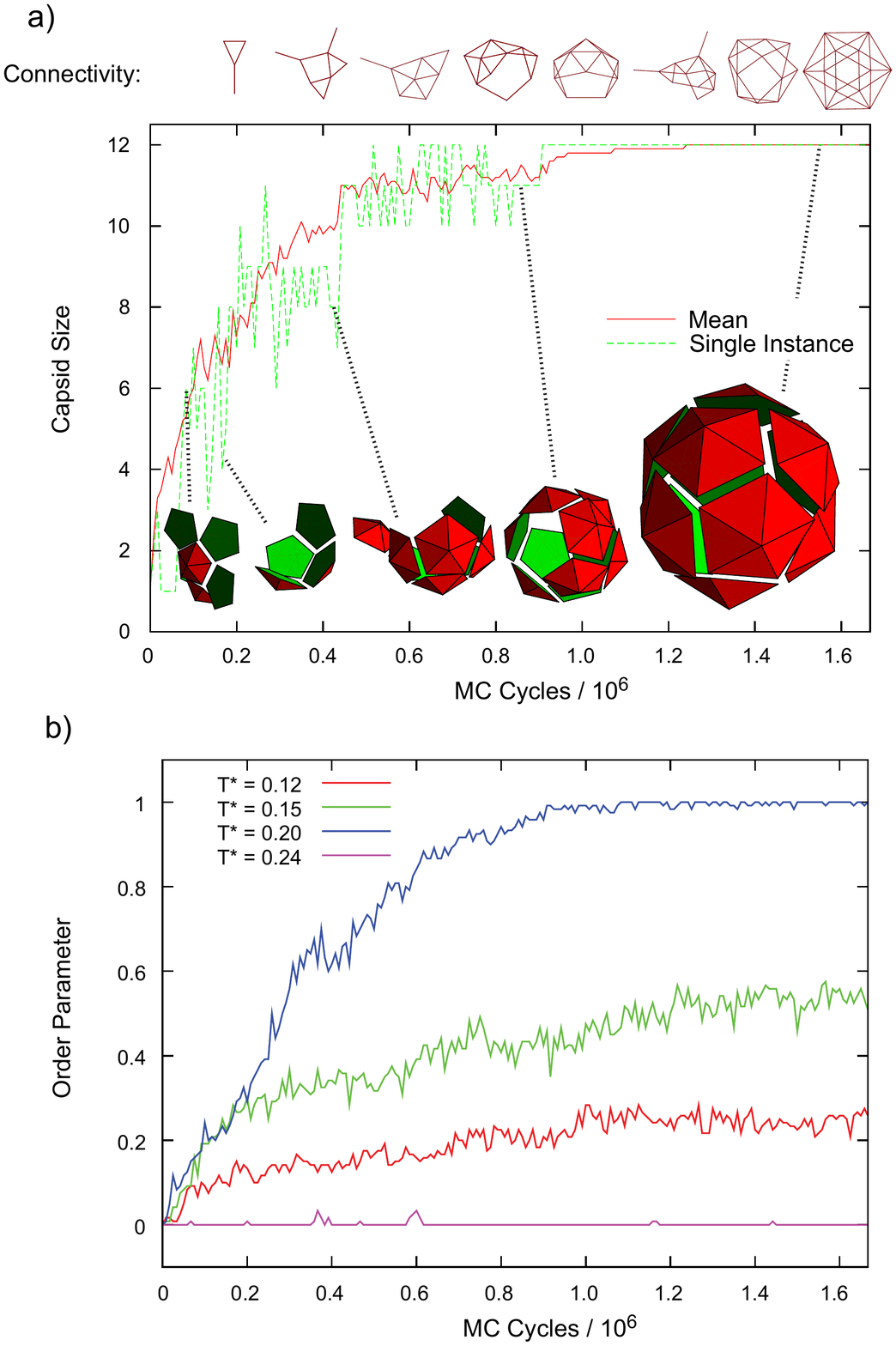}
\end{center}
\caption{ (a) Cluster size order parameter $C$ as a function of simulation time during assembly at $T^* = 0.2\, \epsilon k_B^{-1}, N_5 = 12$ (capable of forming a single capsid).  The cluster size averaged over ten simulations is given by the solid line and a typical single simulation is denoted by the dashed line. Also shown are snapshots and connectivity graphs of the largest cluster at intervals throughout assembly. (b) Geometric order parameter $Q$ with time for different $T^*$.}
\label{fig5}
\end{figure*}

We also monitored the geometry of intermediate structures on the route to full assembly.
Many observed intermediates differ from the most stable structures utilised by rate equation approaches. Here Fig.~\ref{fig5} depicts the connectivity graphs of selected intermediate structures. Nodes represent capsomers in the largest cluster present and edges representing bonds between capsomers. The connectivity graphs of growing clusters often indicate structures with several `leaves' (capsomers with only one bond to the rest of the cluster) whereas the most stable structure for a given size necessarily minimises the number of leaves. In addition, intermediates exist where some capsomers are bonded to a cluster in geometrically incorrect positions, for example, inverted or differently angled. The very commonplace presence of these structures in assembly paths suggests that  the subset used in rate equation studies may omit important detail about the assembly process \cite{nguyen2007deciphering,misra2008pathway,schwartz1998local}.

We also observed hysteresis:  If a capsid was fully formed, and subsequently the temperature is raised, the temperature at which it breaks up is higher than the temperature at which it forms spontaneously from individual capsids.    For simulations with the local MC moves, the difference between the superheated and undercooled temperatures were measured to be  about $\Delta T^* = 0.15\, \epsilon k_B^{-1}$.

\begin{figure}
\begin{center}
\includegraphics[width=8cm]{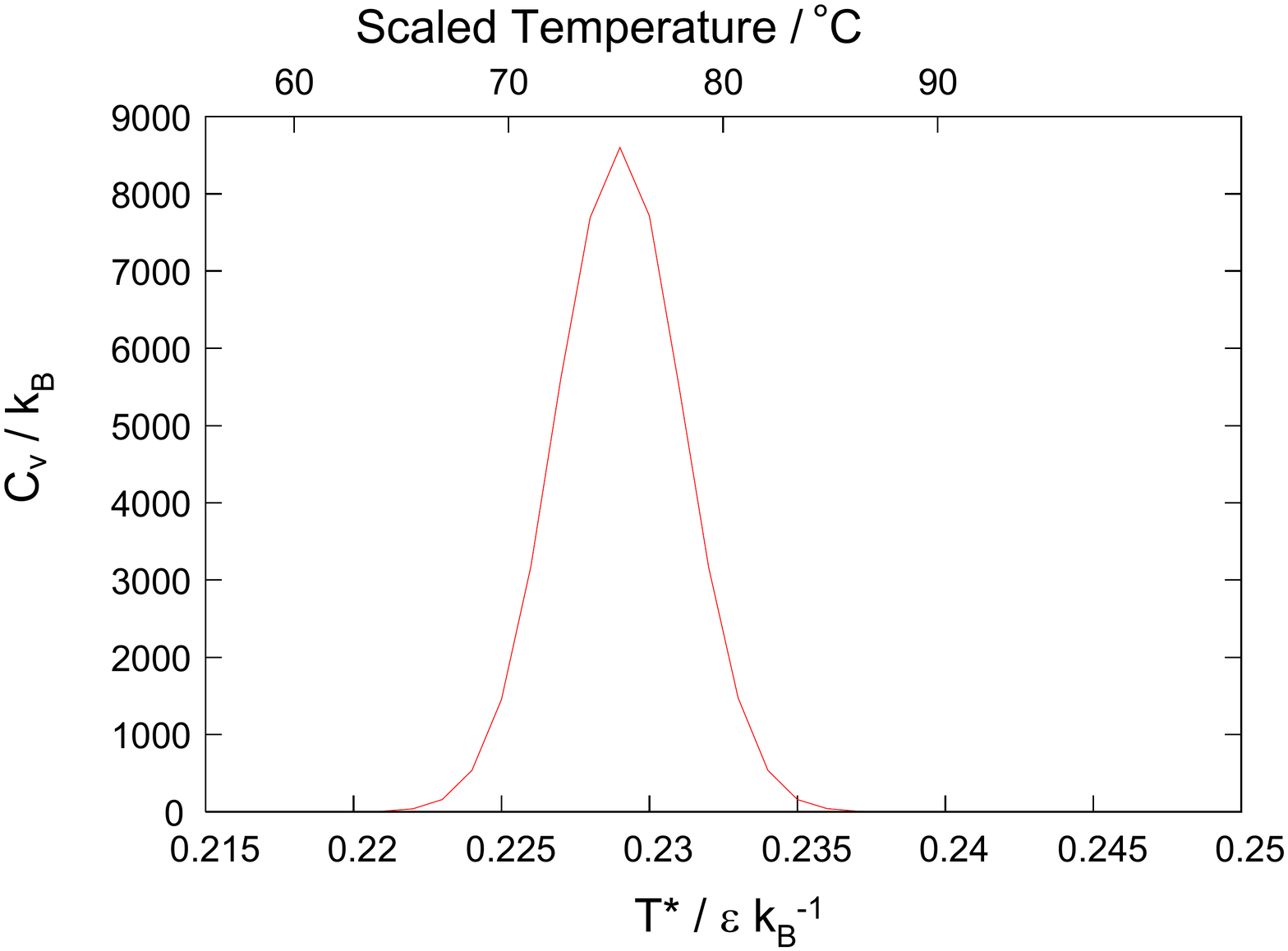}
\end{center}
\caption{Heat capacity from umbrella sampling, $N_5 = 120, \rho = \rho^*, h = 0.5 r_b$. Upper axis shows temperature scaled such that the peak from simulation data matches the peak of an experimental heat capacity curve for a $T = 1$ parvovirus capsid \cite{carreira2004vitro}. Our model displays a full-width at half-maximum (FWHM) on this scale of approximately $7^o$C, rather broader than the experimental FWHM of approximately $2.5^o$C.}  
\label{fig7}
\end{figure}

To fully sample the thermodynamics of the capsid formation, we also performed umbrella sampling MC simulations \cite{torrie1977nonphysical}, with the total energy term adjusted by a term $V'(Q)$ dependent on the order parameter. This allows the system to sample extensively over the full range of $Q$.   From this the melting point  can be determined while the heat capacity $C_v$ follows from   fluctuations in energy.  We observe a peak in $C_v$ at the melting point of the capsid structure (see Fig.~\ref{fig7}), and the area under the peak is  approximately equal to the energy of the fully-formed structure. However, the heat capacity peak is  somewhat  broader than that observed in biological studies \cite{carreira2004vitro, ross2006free}. 

\section{Crowding agents and $T=3$ capsids}

\subsection{Assembly yield with crowding agents}

\begin{figure}
\begin{center}
\includegraphics[width=8cm]{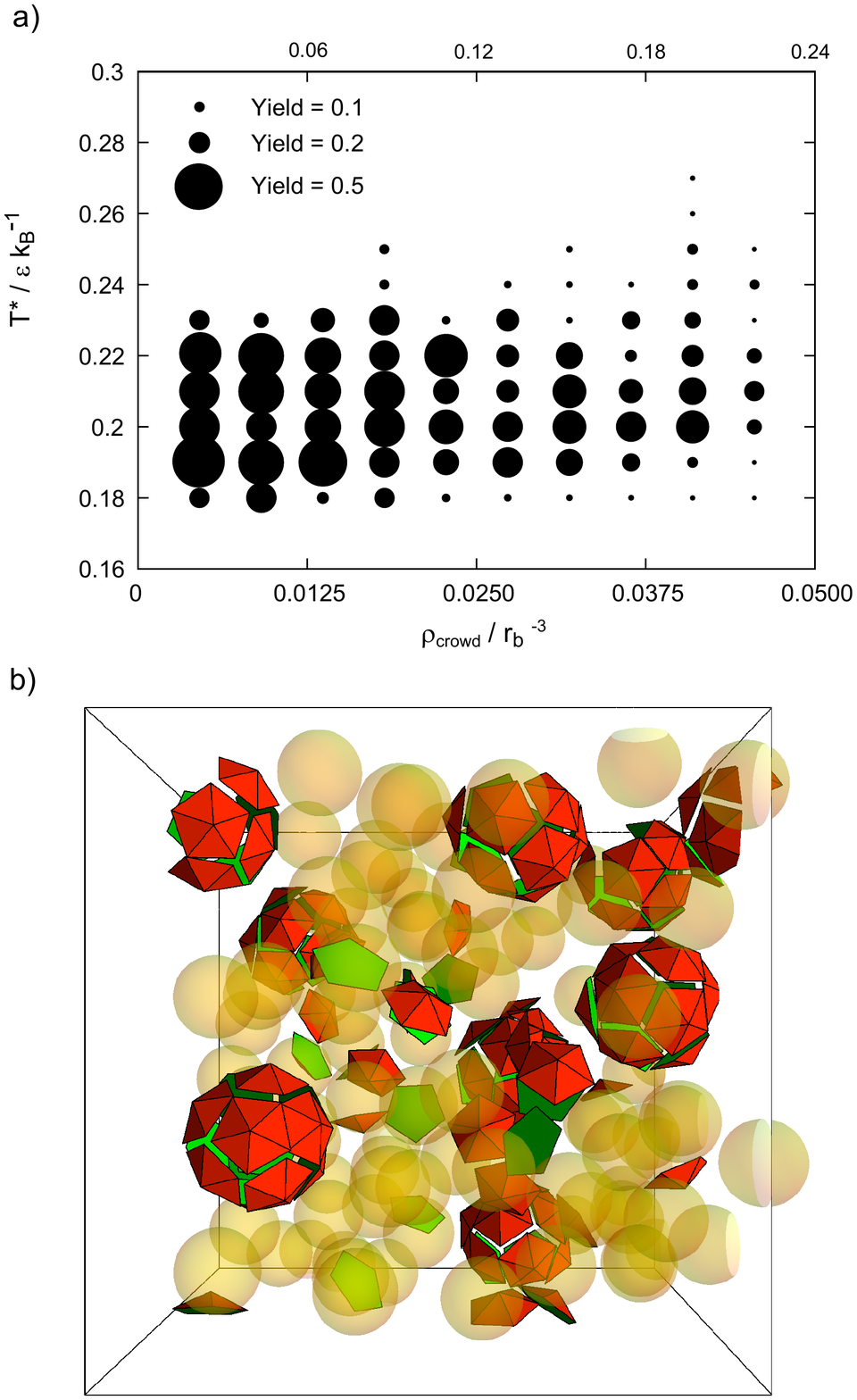}
\end{center}
\caption{(a) Assembly yield with number density $\rho_C$ of crowding agents and temperature $T^*$. (b) Snapshot of assembly with $\rho_c = 4.1 \times 10^{-2}\,r_b^{-3}$ crowding agents (yellow) at $T^* = 0.22 \, \epsilon k_B^{-1}$, showing several fully-formed capsids.}
\label{fig8}
\end{figure}

The focus of this paper has so far been on {\em in vitro} self-assembly of virus capsids.  As a first step towards modeling the biologically more relevant case of {\em in vivo} assembly, we introduced crowding agents into the simulation. In biology, the cytoplasm of a cell is typically filled with a significant volume fraction of proteins and other biomolecules \cite{ellis2003join,fulton1982crowded}, and these are expected to affect the kinetics of assembly. Experimental studies have found the presence of crowding agents to facilitate \emph{in vitro} assembly of HIV-1 capsids \cite{del2005effect} and to stabilise the native states of folded proteins \cite{stagg2007molecular}. We model the crowding by introducing soft repulsive spheres interacting with the potential
\begin{equation}
V_{crowd}(r) = \epsilon \left( \frac{\sigma}{r} \right) ^{12},
\end{equation}
both between their centres and with the apices of the pentagonal pyramids.
Simulations were run with  $N_5 = 120$ capsomers at  $\rho = \rho^*$. 
Assembly yields were measured for different crowding agent densities $\rho_C$ and for different temperatures $T^*$.   The simulation time was lengthened to $1.5 \times 10^6$ MC cycles, three times longer than the bare simulations, because the crowding agents are expected to slow down the diffusion of the capsomers.  Fig.~\ref{fig8} shows the results from these experiments.   Increased crowding  generally lowers the assembly yields.    However, there are exceptions to this.  The crowding was observed to  increase yields at some higher temperatures.  For example, we observe non-zero yields at $T^*=0.27\, \epsilon k_B^{-1}$ at some of the higher crowding densities, while the yield is essentially zero there without crowding agents.
This can be related to Fig.~\ref{densityfig}, where we observe that  the capsomers also show higher yields at higher temperatures for higher densities.  This suggests that the increased yields are due to crowding decreasing the free volume available to the capsomers.

\subsection{Modelling T=3 capsids}

For a triangulation number of $T=1$, $60$ identical proteins group together in $12$ sets of five-fold pentamers.  For higher triangulation numbers, the proteins come together in local six-fold coordination as well.   Although the chemical makeup of the proteins is identical, it is thought that they may undergo allosteric changes that allow them to have five-fold or six-fold bonding arrangements.  Some experiments suggest that capsids can assemble and disassemble from five-fold and  six-fold  subunits \cite{willits2003effects,hanslip2006assembly}.  In that case, it may be a reasonable approximation to treat the assembly as a hierarchical process, where the pentamers and hexamers first form, and then come together into the capsid.

To model larger triangulation numbers, we introduce hexameric particles with the same side lengths as the pentagonal units, which leads to a radius of $1.18\, r_b$.  The hexamer apex has the same height as the pentagon apex, that is $h=0.5\, r_b$, and has the same repulsive interaction. In addition, the interactions between the basal vertices have the same Morse potential form as previously, but with a generalized values of $\epsilon$ given by $\epsilon^{55} = \frac{\epsilon}{2}, \epsilon^{56} = 2 \epsilon, \epsilon^{66} = \epsilon$, where $\epsilon^{ij}$ is the strength of bonding between a vertex on a capsomer with $i$ sides and a vertex on a capsomer with $j$ sides.

We choose these values to discourage pentamer-pentamer contacts. In Ref. \cite{fejer2009energy}, Wales and coworkers compared the PES of 20 hexamers and 12 pentamers for a parameterisation of their original model to that in their newer model, which also includes a selective repulsion site below the plane of the pyramid. For their original model, as well as kinetic traps with inverted capsomers, traps with adjacent pentamers were observed. 

\begin{figure}
\begin{center}
\includegraphics[width=7cm]{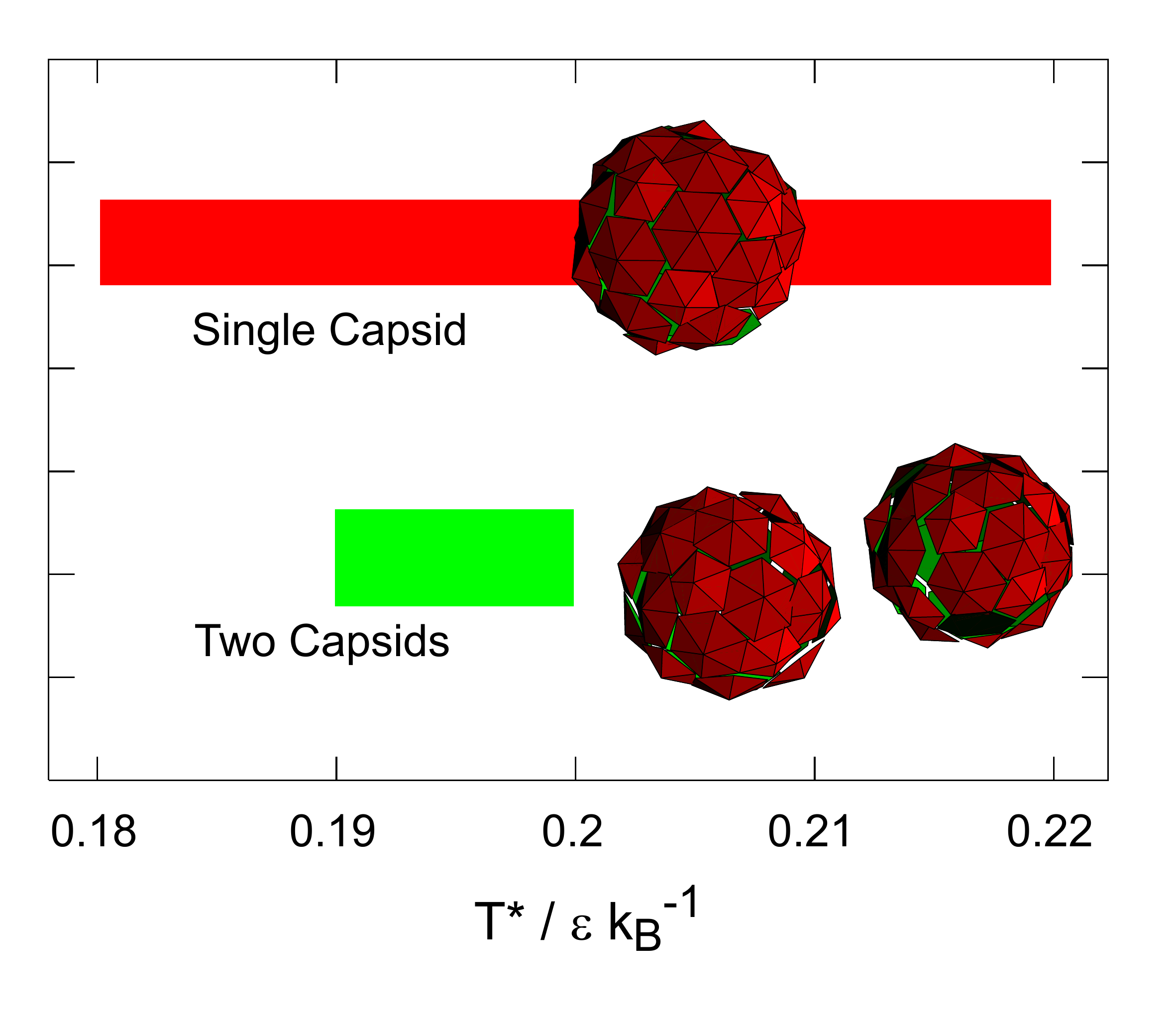}
\end{center}
\caption{Temperature range for successful assembly of single and double $T=3$ capsid structures, with $\rho = \rho^*$ and interaction energies described in the text. The coloured bars represent regions where at least one of ten simulations run displayed correct assembly.}
\label{fig9}
\end{figure}

We  performed similar simulations as before at an effective total capsomer (pentamers and hexamers) density of  $\rho = \rho^*$. 
The  simulations were extended to considerably longer times, up to $5 \times 10^8$ steps, to allow for more complex assembly mechanisms. Assembly of individual $T=3$ capsids was observed across a range of temperatures, but, as shown in Fig.~\ref{fig9}, the assembly of two $T=3$ capsid structures occurred only for a more narrow temperature window.   The lower assembly yields for two capsids are due in part to proto-capsid structures being formed without the correct relative number of pentamer and hexamer units.  In essence, there is an increased mixing entropy term to contend with.   Of course a simulation of just one $T=3$ capsid automatically sets the right ratio of pentamers to hexamers, but it does not guarantee that two separate proto-capsids will not grow, or that pentamers and hexamers will bind to the right positions.

\begin{figure}
\begin{center}
\includegraphics[width=6cm]{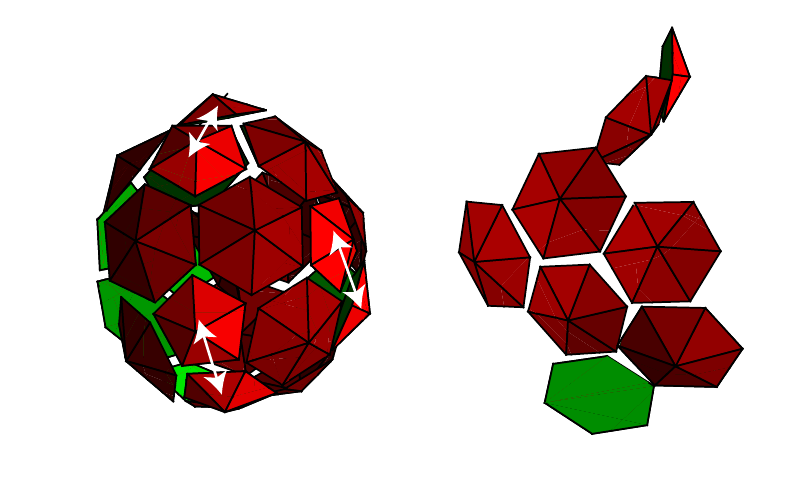}
\caption{Kinetic traps in the formation of a $T=3$ capsid. (left) Proto-capsid with adjacent pentamers (shown with arrow pairs), preventing correct geometric formation. (right) Hexameric sheets.}
\label{fig10}
\end{center}
\end{figure}

Kinetic traps were observed, as in the $T=1$ experiments, with two significant new effects frustrating successful assembly in many cases (see Fig.~\ref{fig10}). Firstly, several proto-capsid structures were observed to form with two pentamers adjacent to one another, fixed in place by pentamer-hexamer bonding. This geometric defect then prevented the further formation of the capsid. Secondly, at lower temperatures, hexamers tend to group into planar sheet structures. A similar phenomenon has been observed in experiments \cite{stasny1968effect},  
where weakening the inter-capsomer bonds of an Adenovirus capsid by treatment with formamide caused the system to initially dissociate into sheet structures.

\begin{figure}
\begin{center}
\includegraphics[width=10cm]{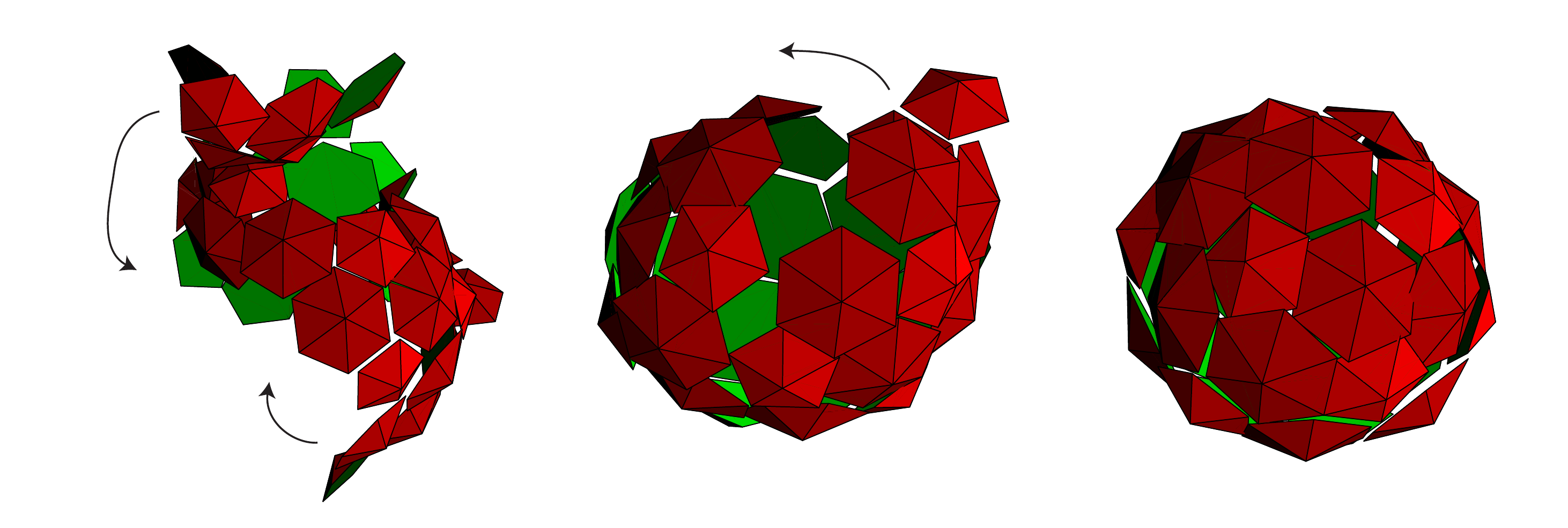}
\end{center}
\caption{Illustration of the `shell rearrangement' mechanism described in the text. (left, $8 
\times 10^6$ cycles) A bonded, misshapen capsid shell has formed. Sections slowly re-orientate to correct geometry. (centre, $20 \times 10^6$ cycles) Proto-capsid now has one major line defect separating two correctly-formed regions. This defect will close, possibly requiring the replacement of some subunits. (right, $400 
\times 10^6$ cycles) The final structure.}
\label{fig11}
\end{figure}

We also observed more varied mechanisms of assembly for a single $T=3$ capsid than what was found for the $T=1$ case of the previous section. For example, we frequently observed large, bonded proto-capsid shells that form quickly in imperfect geometric structures, and then subsequently slowly rearrange to the final icosahedral structure. This is accomplished by the `closing up' of line defects and repositioning of individual subunits, as illustrated in Fig.~\ref{fig11}.

\section{Discussion}
We compare computer simulation results for the self-assembly of a model for  $T=1$ virus capsid to PES predictions made recently by  Wales \cite{wales2005energy}.    In agreement with the PES picture, we find good assembly yields in regimes where the energy landscape shows a ``funnel''-like topology.  Even though the PES calculations were done for a single capsid with all the capsomers connected, the agreement is good, suggesting that this landscape picture may be a fruitful way to analyze how design parameters can aid or hinder  self-assembly.

In addition, our simulations reproduce a number of features seen in other simulations of mono-disperse self-assembly \cite{nguyen2007deciphering,rapaport2008role,elrad2009mechanisms,chen2007simulation,wilber2009monodisperse,nguyen2009invariant,wilber2007reversible,glotzer2004materials,van2006symmetry,glotzer2007anisotropy,wilber2009self,hagan2006dynamic}, such as a range of temperatures and densities that bound the region of successful assembly, sigmoidal assembly dynamics, hysteresis, and a multitude of kinetic traps  including monomer starvation.   We also observe that at typical assembly pathway samples many states that are not the lowest energy for that number of particles.

We further extend our model to include crowding agents, and find that these generally lower the yields compared to the optimum at no crowding, but in some cases can increase yields as well. Finally, we extend the model to include hexameric particles, and study the assembly of $T=3$ capsid structures.   There we find that it is more difficult to assemble multiple capsids because there are now two independent species of particles, and mixing entropy terms play a role.

\vspace{1.5cm}

\bibliographystyle{unsrt}
\bibliography{virus}

\end{document}